\documentstyle[12pt,a4]{article}
\def\pn{\par\noindent}
\def\Win{${\cal W}_{1+\infty}$}
\def\del{\partial}
\def\deb{\bar{\partial}}
\def\da{\dagger}
\def\hb{b^{\da}}
\def\ha{a^{\da}}
\def\BBCT{\,\hbox{\hbox to -2.8pt{\vrule height 6.7pt width .3pt
  \hss}\rm C}}
\def\BBCS{\,\hbox{\hbox to -2.2pt{\vrule height 4.5pt width.2pt
  \hss}$\scriptstyle\rm C$}}
\def\BBCSS{\,\hbox{\hbox to -2pt{\vrule height 3.3pt width .2pt
  \hss}$\scriptscriptstyle \rm C$}}
\def\BBC{{\mathchoice{\BBCT}{\BBCT}{\BBCS}{\BBCSS}}}

\newcommand{\reseteqn}{\setcounter{equation}{0}}
\newcommand{\mysection}{\reseteqn\section}

\begin{document}
\pagestyle{empty}
\begin{raggedleft}
BONN-HE-93-29\\
hep-th/9309083\\
September 1993\\
\end{raggedleft}
$\phantom{x}$\vskip 0.618cm\par
{\huge \begin{center}
Infinite Symmetry in the Fractional Quantum Hall Effect
\end{center}}\par
\vfill
\begin{center}
$\phantom{X}$\\
{\Large Michael Flohr\footnote[1]{Supported by the Deutsche
Forschungsgemeinschaft\\
$\phantom{xxll}$email: flohr@uni-bonn.de, varnhagen@uni-bonn.de}}\\[1ex]
and\\[1ex]
{\Large Raimund Varnhagen\footnotemark[1]}\\[3ex]
{\em Physikalisches Institut\\ der Universit\"at Bonn\\
Nussallee 12\\
D-53115 Bonn\\ Germany\\
%email: flohr@uni-bonn.de, varnhagen@uni-bonn.de
}
\end{center}\par
\vfill
\begin{abstract}
\noindent
We have generalized recent results of Cappelli, Trugenberger and Zemba
on the integer quantum Hall effect
constructing explicitly a ${\cal W}_{1+\infty}$ for the fractional
quantum Hall effect such
that the negative modes annihilate the Laughlin wave functions. This
generalization has a nice interpretation in Jain's composite fermion theory.
Furthermore, for these models we have calculated the wave functions of the
edge excitations viewing them as area preserving deformations of an
incompressible quantum droplet, and have shown that the
${\cal W}_{1+\infty}$ is the underlying symmetry of the edge excitations
in thefractional quantum Hall effect. Finally, we have applied
this method to more general wave functions.
\end{abstract}
\vfill
%email: flohr@uni-bonn.de, varnhagen@uni-bonn.de
\newpage
\setcounter{page}{1}
\pagestyle{plain}
\mysection{Introduction}
\pn
The experimental discoveries of the integer quantum Hall
effect (IQHE) \cite{Kli80}
and of the fractional quantum Hall effect (FQHE) \cite{TSG82}
are one of the most interesting physical phenomena in solid
state physics in recent years. The conductance of a two dimensional
electron gas in a high magnetic field at low temperature exhibts
quantized plateau values of the form $\sigma_{xy}={e^2 \over h}\nu$,
where the filling factor $\nu$ is an integer or fractional number.
In many respects, both the integer as well as the fractional
effect share very similar underlying physical
characteristics and concepts, for instance the two-dimensionality
of the system, the quantization of the Hall conductance with
simultaneous vanishing of the longitudinal resistance, and the
interplay between disorder and the magnetic field giving rise to the
existence of extended states. In other respects, they encompass
entirely different physical principles and ideas. In particular,
while the IQHE is thought of essentially as a noninteracting electron
phenomenon \cite{Lau81}, the FQHE is believed to arise
from a condensation of the
two-dimensional electrons into a new incompressible state of matter as a
result of interelectron interaction \cite{Lau83}, see also \cite{PG87,Sto92}.
\pn
An important step was taken by Laughlin writing down the wave functions
for the
fundamental fractions $\nu = {1 \over 3},{1 \over 5},{1 \over 7} \dots$
which played a special role
in a hierarchial scheme in which a daughterstate was
obtained at each step from a condensation of quasiparticles of the
parent state into a correlated low-energy state \cite{Hld83,Hlp84}.
Extensive calculations have proven these wave functions to be extremely
close to the numerical exact solutions \cite{PG87}.
\pn
In the last years J.K.~Jain developed the composite
fermion theory which could describe IQHE and FQHE by a common principle
attaching to each electron an even number of magnetic flux quanta
which gives an easy explanation of the experimentally observed
fractional fillings as well as a new derivation of Laughlin's wave functions
starting from the well understood IQHE \cite{Jai89}. New experiments are
in good agreement with this theory \cite{DST93,WRP93,HLR93}.\pn
The incompressibility of these quantum fluids is explained by an finite
energy gap above the ground state. Recently, for the IQHE ($\nu =1$) it was
shown that incompressibility also results in an infinite symmetry which
describes the area preserving nonsingular deformations of the
quantum droplet and commutes with the hamiltonian \cite{CTZ93}.
The quantization of this symmetry is well
known in physics as the nonsingular part of a \Win\ and arises e.g.\ in
string theories or two dimensional gravity \cite{PRS90,Bak89,Wit92}.
These deformations
are directly related to edge excitations which should live on the
one dimensional boundaries and were studied by a number of authors
\cite{Hlp82,Wen92,BW90,FK91,FZ91,CDT93}.
The dynamics of these edge states is mainly based on the relation
of Chern-Simons gauge theories and conformal field theory
\cite{Wit89}. \pn
In this paper we give a generalization of this infinite symmetry to the
FQHE ($\nu = {1 \over 2p+1}$) showing that the Laughlin wave functions
are annihilated by the negative, nonsingular generators of the \Win.
The very interesting point is that, constructing the \Win\ for the
FQHE, interelectron interaction effects enter which automatically cancel
out for the IQHE and agree with the result of reference \cite{CTZ93}.
Furthermore,
we can show that these interactions can be interpreted as arising from
an even number of magnetic flux quanta which are attached to the
electrons as in the composite fermion picture. It turns out that the
interaction is hidden in a nontrivial measure which is the $N$-point
function of $N$ localized flux quanta vortices and should be described
by an abelian Chern-Simons theory.
\pn
Viewing the
QHE-states as a droplet of an incompressible quantum fluid, the
gapless edge excitations can be interpreted coming from surface
waves or area preserving deformations of the droplet.
We have calculated the wave functions for edge excitations with
$\nu={1 \over 2p+1}$ using the fact that they are generated by the
positive modes of the \Win. Here our result agrees with former ones of
M.~Stone for $\nu=1$ and X.G.~Wen for the FQHE \cite{Sto91,Wen92}.
\pn
Finally, we apply the previous method to more general wave functions
describing multi layer systems or systems of interacting Landau levels
for every fractional filling and show that the \Win\ is indeed the
fundamental symmetry of the edge excitations.
\pn
The paper is organized as follows: First we give an introduction to the
basics of the QHE. Next we show how to generalize the construction of
the \Win\ from the IQHE to the FQHE and interprete this generalization
by Jain's composite fermion theory. Then we calculate the wave functions
of the edge excitations using the \Win. Finally, we consider the case of
more general wave functions.

\mysection{Preliminaries}
Let us start by reviewing some elementary facts about a two dimensional
electron in an uniform, transverse magnetic field $B$.
The Schr\"odinger equation for such an electron is given by
\begin{equation}
    H\psi = {1\over 2m}({\bf p} - {e \over c} {\bf A})^2\psi = E\psi\,,
\end{equation}
where the momentum ${\bf p} = -i\hbar{\bf \nabla}$ and the
gauge potential
${\bf A}$ live in the plane. This problem can be solved exactly.
Let us choose the symmetric gauge
${\bf A}={B \over 2}(-y,x)$ and introduce complex variables:
$z=x+iy$, $\bar z = x-iy$ and $\del = {1\over 2}(\del_x - i \del_y)$,
$\deb = {1\over 2}(\del_x + i\del_y)$. Defining all lengths in units of the
magnetic length,
\begin{equation}
   l=\Big({2\hbar c\over eB}\Big)^{1\over 2}\,,
\end{equation}
and the energies in units of the Landau level spacing,
\begin{equation}
   \omega_c = {eB \over mc}\,,
\end{equation}
the Hamiltonian can be reexpressed as:
\begin{equation}
  H = 2\hbar\omega_c l^{2}\big(-\del\deb + {1\over 2l^2}(\bar z\deb - z\del)
      + {1\over 4l^4}z\bar z \big) \,.
\end{equation}
Letting $\hbar=m=l=1$ the hamiltonian and the angular momentum $J$ can
be written
in terms of a pair of independent harmonic oscillators:
\begin{eqnarray}
   H &=& a^{\da}a + aa^{\da}\,,\\
   J &=& b^{\da}b - a^{\da}a\,,
\end{eqnarray}
where these operators are
\begin{eqnarray}
    a={z \over 2} + \deb\,,& \qquad\qquad &a^{\da} = {\bar z \over 2}
    - \del\,,\\
    b={\bar z \over 2} + \del\,, & \qquad\qquad &b^{\da} = {z \over 2}
    - \deb\,,
\end{eqnarray}
and satisfy the commutation relations
\begin{equation}
   [a,a^{\da}] = 1\,, \qquad\qquad [b,b^{\da}] = 1\,,
\end{equation}
with all other commutators vanishing. The vacuum is determined by the
condition $a\psi_{0,0} = b\psi_{0,0} = 0$ and given as
\begin{equation}
   \psi_{0,0} = {1 \over  \sqrt{\pi}}\exp(-{1 \over 2}\mid z \mid^2)\,.
\end{equation}
In terms of the operators $a^{\da}$
and $b^{\da}$ the solutions can finally be written as
\begin{equation}
   \psi_{n,l} = {(\hb)^l (\ha)^n \over \sqrt{l!n!}}\psi_{0,0}
\end{equation}
with energy $E_n = 2n + 1$, which determines the Landau level.
These energy states are infinitely
degenerate due to  the rotational invariance around the z-axis.
It is usefull to note that in the lowest Landau level the polynomial part
of the  wave function is holomorphic, and involves in the second
Landau level
at most one power of $\bar z$. In general the highest power of
$\bar z$ in the $n^{th}$-Landau level is $n-1$. \pn
In a finite sample of area $A$
one can show, that the degeneracy of each Landau level
is determined by the number of the magnetic flux quanta
\begin{equation}
   N_A = {\Phi_{mag}\over \Phi_0}\,,
\end{equation}
where $\Phi_{mag}=BA$ is the magnetic flux through the
area $A$ and $\Phi_0={h \over e}$ is a single flux quantum.
\pn
Let us now consider the case of $N$ such electrons. If there is no
interaction between them, the many particle problem splits into
$N$ copies of the single particle problem. Therefore, we just get $N$
operators, identical to the single particle operators $a,b$, but now
labelled by an index $i$ refering to the coordinate of the $i^{\rm th}$
electron: $a_i, b_i$.
Since the magnetic field $B$ controls the number of states and thus the
density of electrons per state, its action can be considered as an external
pressure. Actually, the electron density per state is the correct quantum
measure of the electron density, the filling fraction $\nu$
\begin{equation}
        \nu = {N \over N_A} \,.
\end{equation}
The IQHE is well understood by a gauge argument of Laughlin. Later
it was shown that the conductivity can be interpreted as the Chern
character of a $U(1)$-fibre bundle over a torus \cite{NTW85,Koh85,AS85}
or as an element in the cyclic cohomology of a
$C^{\star}$-algebra \cite{Bel86}.
\pn
For the FQHE with filling fraction $\nu = {1 \over 2p+1}$ by numerical
experiments Laughlin found the groundstates given by the following
wave functions:
\begin{equation}
   \psi_p = \prod_{i < j} (z_i - z_j)^{2p+1}\exp(-{1\over 2}\sum_i
            \mid z_i \mid^2)\,,
\end{equation} where $p$ should be an integer to respect the Pauli
principle.
In the composite fermion theory this wave function was reinterpreted
by J.K.~Jain as a wave function not of bare single electrons,
but of
electrons bound to an even (here $2p$) number of vortices or flux
quanta. Starting with the wave function $\phi_n$ of the IQHE with filling
fraction $\nu = n$ one attaches $2p$ flux quanta to each electron, which is
given by multiplying $\phi_n$ with $D^{2p}$,
\begin{equation}
    \psi_{\nu} = D^{2p}\phi_{n}\ \ {\rm with}\ \ D = \prod_{i<j}
    (z_i - z_j) \,.
\end{equation}
Using mean field arguments,
this leads to an electron state in which $n^{-1} \pm 2p$ flux quanta are
available to each electron. Thus this composite fermion state has
filling fraction \cite{Jai89}
\begin{equation}
  \nu = {n \over 2pn\pm 1}\,.
\end{equation}
Thus, the Laughlin wave functions are given for $n=1$.
When calculating some expectation values via path integrals, only
closed paths contribute to the partition function, because it is the
trace of $\exp(-\beta H)$. Closed paths are given by exchanging electrons
or by moving them around each other. The phase associated with each path has
two contributions. One is the statistical phase due to the Fermi
statistics of the electrons, and the other one is the Aharanov-Bohm phase
due to the flux enclosed in the loop. But adding to a fermion an even
number of flux quanta again gives a fermion and also the Aharanov-Bohm phase
factor is the same because a flux quantum produces a phase factor of
unity. Thus, adding an even number of flux quanta to each electron does
not change the expectation values. This argument is due to A.~Lopez and
E.~Fradkin \cite{FL}.

%\mysection{{\boldmath\Win}\ for {\boldmath$\nu=1/m$}}
\mysection{\Win\ for $\nu=1/m$}
The wave functions of the last section should describe the condensation
of the electrons to
new states of matter, to incompressible quantum superfluids. Normally,
the incompressibility is explained by a finite energy gap above the
ground state.
Recently A.~Cappelli, C.A.~Trugenberger and G.R.~Zemba have given
another explanation
of this incompressibility for the $\nu=1$ case; they have
found a \Win\ symmetry which is the algebra of the area preserving
nonsingular
diffeomorphisms commuting with the hamiltonian of the system, defining
an incompressible state now to be a highest weight vector of the \Win\
\cite{CTZ93}.
They constructed the generators of the \Win\ in the following way:
\begin{equation}
   {\cal L}_{m,n} = \sum_{i=1}^N (b_i^{\da})^{m+1} (b_i)^{n+1}
   \qquad {\rm for}\quad n,m \geq -1      \,.
\end{equation}
These generators communte with the hamiltonian of the system
and fulfill the following commutation relations:
\begin{equation}
   [{\cal L}_{n,m},{\cal L}_{k,l}] = \sum_{s=0}^{{\rm Min}(m,k)}
         {(m+1)!(k+1)! \over (m-s)!(k-s)!(s+1)!}
         {\cal L}_{n+k-s,m+l-s} - (m \leftrightarrow l,
          n \leftrightarrow k) \,.
\end{equation}
Then they have shown that
\begin{equation}
    {\cal L}_{m,n} \psi_0 = 0 \qquad {\rm for} \qquad n > m \geq -1\,,
\end{equation}
which means that $\psi_0$ is a highest weight vector of the algebra
of area preserving nonsingular diffeomorphisms.
\pn
The aim of our paper is to generalize this result to the FQHE.
We are doing this by changing the definition of teh $b_i$
introducing an interaction term in the following way:
($b_i^{\da}$ remains unchanged)
\begin{equation}
   b_i = \del_i + {\bar z_i \over 2} - 2p \sum_{i \neq j}
   {1\over z_i - z_j}\,.
\end{equation}
For $p=0$ one recovers the original definition for the $b_i$ and
$b^{\da}_i$
as before, so we are not changing our notation. The commutators of
the $b_i$ and $b_i^{\da}$ change in the following way
\begin{eqnarray}
     [b_i,b^{\da}_i] &=& 1 + 2p\pi \sum_{i\neq j} \delta(z_i - z_j)\,,\\
   \ [b_i,b^{\da}_j] &=& -2p\pi \delta(z_i - z_j)\ \ {\rm for}\ \ i\neq j\,.
\end{eqnarray}
%The occurrence of the delta functions will be explained below.
Defining the ${\cal L}_{mn}$ as above but with the
new $b_i$ they fulfill the commutation relations of the
same \Win\ up to terms involving delta functions. In the case of
fermions, which have to respect the Pauli principle, the delta functions
do not contribute, since the wave function has to approach zero
for $z_i \longrightarrow z_j, i\neq j$. For the first Landau level,
where the wave functions are holomorphic up to the exponential term,
one can rewrite the operators $b_i$ and $b_i^{\da}$ such that they only
act on the holomorphic part,
\begin{eqnarray}
    b_i&=&\del_i - 2p \sum_{i \neq j}{1\over z_i - z_j}\,,\\
    b_i^{\da}&=&z_i\,.
\end{eqnarray}
Note, that $b_i^{\da}$ just acts by multiplication. Thus, in the case of
the first Landau level no delta functions will occur.
In the standard notation of \Win\ we set
\begin{equation}
   W^{(s)}_n \sim {\cal L}_{n+s-2,s-2}\,,\qquad s \geq 1\,,\qquad
   n \geq -s+1\,,
\end{equation}
where $W^{(s)}_n$ is the $n^{th}$-Fourier mode of a spin $s$ field.
After some calculations which can be found in the appendix
one obtains the action of the modes $W^{(s)}_n$ on the Laughlin wave
function $\psi_p$:
\newfont{\bigrm}{cmr17 scaled 3583}
\begin{equation}\label{eq:mammut}
  W^{(s)}_n \psi_p = (s-1)!%\raisebox{-2ex}{
  %$\displaystyle
  \sum_{1 \leq j_0 < j_1 < \dots < j_{s-1} \leq N}%$}
  \frac{
    \phantom{l}\left|
      \begin{array}{ccccc}
        1 & 1 & \cdots & 1 & 1 \\
        z_{j_0} & z_{j_1} & \cdots & z_{j_{s-2}} & z_{j_{s-1}} \\
        \vdots  & \vdots & \ddots & \vdots & \vdots \\
        z_{j_0}^{s-2} & z_{j_1}^{s-2} & \cdots & z_{j_{s-2}}^{s-2} &
        z_{j_{s-1}}^{s-2} \\
        z_{j_0}^{s-1+n} & z_{j_1}^{s-1+n} & \cdots & z_{j_{s-2}}^{s-1+n} &
        z_{j_{s-1}}^{s-1+n}
      \end{array}
    \right|\raisebox{-8ex}{\phantom{l}}
  }{
    \raisebox{8ex}{\phantom{l}}\left|
      \begin{array}{ccccc}
        1 & 1 & \cdots & 1 & 1 \\
        z_{j_0} & z_{j_1} & \cdots & z_{j_{s-2}} & z_{j_{s-1}} \\
        \vdots & \vdots & \ddots & \vdots & \vdots \\
        z_{j_0}^{s-2} & z_{j_1}^{s-2} & \cdots & z_{j_{s-2}}^{s-2} &
        z_{j_{s-1}}^{s-2} \\
        z_{j_0}^{s-1} & z_{j_1}^{s-1} & \cdots & z_{j_{s-2}}^{s-1} &
        z_{j_{s-1}}^{s-1}
      \end{array}
    \right|\phantom{l}
  }\,\psi_p
\end{equation}
from which immediately follows that acting on $\psi_p$ the negative modes
vanish,
\begin{equation}
    W^{(s)}_n \psi_p = 0 \qquad {\rm for} \qquad -s < n \leq -1\,.
\end{equation}
Moreover, the states $\psi_p$ are eigenstates for the zero modes,
\begin{equation}
    W^{(s)}_0 \psi_p = (s-1)!{N \choose s} \psi_p\,.
\end{equation}
Let us emphasize this result:
We have shown, that the Laughlin wave functions are highest weight states
of the quantized algebra of non singular area preserving diffeomorphisms
which means that all surface waves on the droplet move in the same
direction. The
singular deformations cannot be included in the algebra in that way,
since they would change the topology of the droplet.
Now, we have a common formulation for $\nu=1$
and $\nu={1\over 2p+1}$ QHE, where automatically the IQHE is described by
an single electron theory, but the FQHE needs interelectron interaction
in the first Landau level. \pn
At that point the reader may worry that $b_i$ and
$b_i^{\da}$ are not hermitian conjugate. However, we can take an inner
product of the form
\begin{equation}
         <\Psi_1 \mid \Psi_2> = \int \Psi_1^{\da} \mu \Psi_2^{}\,,
\end{equation}
where $\mu$ is given as:
\begin{equation}\label{eq:mu}
    \mu(z_1,\bar z_1,\dots,z_N,\bar z_N) = \prod_{i<j}
    \mid z_i - z_j \mid^{-4p}\,.
\end{equation}
Using this inner product $b_i$ and $b_i^{\da}$ become hermitian conjugate
to each other.
One will see that this measure is very important in the following,
especially for the interpretation of the new interaction term in the
$b_i$s. Namely introducing the nontrivial measure the hamiltonian
would be non hermitian. Thus we have to change the definition of the
$a_i^{\da}$ in the following way: ($a_i$ remains also unchanged)
\begin{equation}
   a_i^{\da} = -\del_i + {\bar z_i \over 2} + 2p  \sum_{i \neq j}
   {1\over z_i - z_j}
%   \qquad {\rm and} \qquad
%   a_i = \deb_i + {z_i \over 2}
   \,.
\end{equation}
The commutation relations are now given as:
\begin{eqnarray}
     [a_i,a_i^{\da}] &=& 1 - 2p\pi \sum_{i\neq j} \delta(z_i - z_j)\,,\\
   \ [a_i,a_j^{\da}] &=& 2p\pi \delta(z_i - z_j)\ \ {\rm for}\ \ i\neq j\,.
\end{eqnarray}
The hamiltonian is defined as before
\begin{equation}
    H = \sum_{i=1}^N(\ha_i a_i + a_i \ha_i)
\end{equation}
and commutes with the \Win\ without occurence of any delta functions.
The Landau level structure is not destroyed
and the Laughlin wave function for $\nu = {1 \over 2p+1}$ is an
eigenfunction in the lowest Landau level. \pn
The configuration space for distinguishable particles is given by
\begin{equation}
    {\rm C}_N = \{(z_1,\dots,z_N) \in \BBC^N; \quad z_i \neq z_j
            \quad {\rm for} \quad i \neq j \} \,.
\end{equation}
The $(a_i,\ha_i)$ can be considered as covariant derivatives on a $U(1)
\otimes \dots \otimes U(1)$ bundle over C$_N$ as in the paper
of E.~Verlinde on the non-abelian Aharanov-Bohm effect \cite{Ver91}.
Thus, the curvature is given
by (3.14) which describes a constant magnetic field plus $2p$ flux
quanta added to each electron. This is exactly the FQHE interpretation of
J.K.~Jain by the composite fermion theory mentioned
in the previous section. These flux quanta can be described in an
abelian Chern-Simons theory by localized Wilson loops. Considering
the $N$-point function of these flux quanta localized at the positions $z_i$
of the electrons one sees that it is
proportional to the measure $\mu$ (\ref{eq:mu}) using that those Wilson
loop operators can be expressed by vertex operators \cite{BBGS92}.
This explains
the former observation on the relation between vertex operator
correlators and the Laughlin wave function \cite{Fub91,Sto91,CMM91,
MR91}. \pn
This picture is in good agreement with the argument of Lopez
and Fradkin stated previously that adding an even number of flux quanta to
each electron leaves all expectation values invariant.
Calculating the expectation values of the Laughlin wave function one
also has to introduce the measure $\mu$ (\ref{eq:mu}):
\begin{equation}
   \int \psi_p^{\da}\,\mu\, \psi_p^{\phantom{\da}} dz^N \,.
\end{equation}
It is easy to see that this expression is independent of $p$, thus, adding
flux quanta does not change the expectation value.
\pn
Thus, in our formulation of
the FQHE we consider a hamiltonian without explicit
interelectron interaction as in the IQHE,
but describing the interaction with the help of a nontrivial
measure coming from the $N$-point correlation function of the flux
quanta in an abelian Chern-Simons theory.

\mysection{Edge excitations}
Halperin was the first who pointed out that the IQHE states contain
gapless edge excitations, which are responsible
for nontrivial transport properties \cite{Hlp82}.
Using gauge arguments, one can easily show
that FQHE states also support gapless edge excitations.
X.G.~Wen has shown that these states span a representation
of a Kac-Moody current algebra \cite{Wen92} and M.~Stone has described
them using Schur functions or homogenous symmetric polynomials \cite{Sto91}.
In this section
we derive these results with the help of the \Win.
\pn
Viewing the QHE-states as a droplet of an incompressible quantum fluid
we consider the edge excitations as area preserving deformations of the
droplet, which are described by the \Win. Thus, the highest weight
representation on the QHE wave function should give the spectrum
of these edge states:
\begin{equation}\label{eq:www}
  W^{(s_1)}_{n_1}W^{(s_2)}_{n_2} \dots W^{(s_k)}_{n_k}\psi_p
  \,,\ s_i\geq s_{i+1}\,.\
  n_i\geq n_{i+1}\ {\rm if}\ s_i=s_{i+1}\,,
\end{equation}
In fact, equation (\ref{eq:mammut}) shows, that applying one mode of a \Win\
generator to $\psi_p$ yields $\psi_p$ multiplied by a symmetric function,
since the fraction of the determinants is equal to the Schur function
${\cal S}^{\{0,0,\ldots,0,n\}}$. Actually, every Schur function can be
written as a fraction of a certain determinant and the Vandermonde
determinant. If we use the notation
\begin{equation}
  D^{\{m_1,m_2,\ldots,m_n\}} = \left|\begin{array}{cccc}
  z_1^{m_1}    & z_2^{m_1}    & \ldots    & z_n^{m_1}    \\
  z_1^{1+m_2}  & z_2^{1+m_2}  & \ldots    & z_n^{1+m_2}  \\
  \vdots       & \vdots       & \ddots    & \vdots       \\
  z_1^{n-1+m_n}& z_2^{n-1+m_n}& \ldots    & z_n^{n-1+m_n}
  \end{array}\right|\,,
\end{equation}
then the Schur functions can be expressed as \cite{Sto91}
\begin{equation}
  {\cal S}^{\{m_1,m_2,\ldots,m_n\}} =
  \frac{D^{\{m_1,m_2,\ldots,m_n\}}}{D^{\{0,0,\ldots,0\}}}\,.
\end{equation}
By induction it follows that monomials as in
(\ref{eq:www}) also yield polynomial symmetric functions multiplied with
$\psi_p$. The reason is that any such state must be totally antisymmetric due
to the Pauli principle. But since $\psi_p$ is always reproduced, the only way
to get a totally antisymmetric polynomial function is to multiply $\psi_p$ by
a totally symmetric one.
Moreover, the current $j \equiv W^{(1)}$ already
yields a complete set of symmetric functions, namely the products of
power sums $s_{k} = \sum_{i=1}^{N} z_i^k$,
\begin{equation}
  j_{n_1}j_{n_2}\ldots j_{n_k}\psi_p = s_{n_1}s_{n_2}\ldots s_{n_k}\psi_p\,.
\end{equation}
This is a basis of all symmetric functions provided $n_i\geq n_{i+1}$.
To see this, one just has to note that the action of
\begin{equation}
  W^{(1)}_k = \sum_{i=1}^{N}(b^{\da}_i)^k
\end{equation}
on $f(z_1,\ldots,z_N)\exp(-\frac{1}{2}\sum_{i=1}^{N}|z_i|^2)$ with
$f(z_1,\ldots,z_N)$ any holomorphic function on $\BBC^N$
is just given by the multiplication
with $s_{k}$ which is a holomorphic function for $k \geq 0$.
\pn
Thus, the \Win\ algebra yields all possible edge excitations which respect the
Pauli principle. The resulting
spectrum is given by the set of all symmetric polynomial functions with
the partition function being nothing but
\begin{equation}\label{eq:part}
  Z(q=e^{2\pi i\tau}) = \sum_{n=0}^{\infty}p(n)q^n =
  \prod_{n=1}^{\infty}{1\over 1 - q^n}\,,
\end{equation}
where $p(n)$ denotes the number of partitions of $n$ in positive integers.
Thus, the positive modes of the current $j \equiv W^{(1)}$ alone generate
all edge excitations which means that these excitations can be interpreted
as surface waves moving in the same direction and moving with the same
velocity. Therefore the spectrum
is equivalent with that of the $U(1)$-Kac-Moody algebra at level 1. In this way
the results of X.G.~Wen and M.~Stone reappear in an unified way
\cite{Wen92,Sto91}.
\pn
These considerations show that the conformal field theory which corresponds
to the Chern Simons theory describing the attachment of flux
quanta to the electrons and which is defined on the boundary of the
system (the Laughlin droplet) must be generated by a $U(1)$-Kac-Moody
current. Thus, it follows that the conformal theory must have the
effective central charge $c_{{\rm eff}} = 1$. This aggrees with the fact
that the non-trivial measure introduced in the third section, where it
arose from the Knizhnik-Zamolodchikov connection describing the effect
of the attached flux quanta, is given by a correlation function of a
$c_{{\rm eff}} = 1$ conformal field theory.

\mysection{Generalizations}
\pn
There exit a lot of other examples of trial wavefunctions not only for
filling fraction $\nu={1 \over 2p+1}$.Most of these wavefunctions have
the following structure \cite{Wen92,FZ91,WZ92,Jai89a}:
\begin{equation}
\psi_K = \prod_{I<J} \prod_{i\leq j}(z_i^I - z_j^J)^{K_{I,J}}\prod_I
         \prod_{i<j}(z_i^I - z_j^I)^{K_{I,I}}
         \exp(-{1\over 2}(\sum_{i,I} \mid\! z_i^I \! \mid^2)\,,
\end{equation}
where $K$ is a symmetric, integer valued $m \times m$ Matrix with odd integers
on the main diagonal. Then, the filling fraction is given by
\begin{equation}
   \nu = \sum_{I,J} (K^{-1})_{I,J}\,.
\end{equation}
Thus, one can get different wave functions for the same filling fraction
$\nu$.
The physical picture behind this ansatz is to couple different
independent Hall fluids (i.e.\ sets of eventually interacting Landau
levels or different layers).
Viewing the filling fraction $\nu$ to be proportional to the Hall
conductivity $\sigma$ one sees, that the total Hall conductivity is
determined by the Hall conductivities of the several Hall fluids (or
Landau levels) according to the Kirchhoff rules for coupling them in
parallel or in series.
\pn
Now, it is easy to see that the \Win\ can be constructed in the same way
as before defining $b_i^I$ and ${b_i^{I \da}}$ as
\begin{eqnarray}
   b_i^I &=& \del_i^I + {{\bar z}_i^I \over 2} + \sum_{J\neq I}A_{I,J}
             \sum_{i\leq j}{1 \over z_i^I - z_j^J} + A_{I,I}
             \sum_{i<j}{1 \over z_i^I - z_j^I}\,,\\
   b_i^{I\da} &=& -\deb_i^I + {z_i^I \over 2}\,,
\end{eqnarray}
and
\begin{equation}
   {\cal L}_{m,n} = \sum_{I,i}(b_i^{I \da})^{m+1}(b_i^I)^{n+1}\,.
\end{equation}
The heighest weight condition can be fulfilled, if
\begin{equation}
   K = 1\!{\rm l} - A\,,
\end{equation}
where $1\!$l is the $m \times m$ identity matrix.
For example, the $\nu = \frac{m}{2pm+1}$ FQHE can be obtained if $K$ is
given by the following $m \times m$ matrix \cite{FZ91,WZ92},
\begin{equation}
   K = \left(\begin{array}{cccc}
       2p+1 & 2p   & \ldots & 2p \\
       2p   & 2p+1 &        & 2p \\
       \vdots & &\ddots & \vdots \\
       2p & \ldots & 2p & 2p+1
       \end{array}\right)\,.
\end{equation}
Thus, the matrix $A$ is nothing but
\begin{equation}
   A = -2p\left(\begin{array}{ccc}
                 1 & \ldots & 1 \\
            \vdots & \ddots & \vdots \\
                 1 & \ldots & 1
          \end{array}\right)\,,
\end{equation}
which indeed can be considered as the addition of $2p$ magnetic flux
quanta to each particle as stated previously.
\pn
The edge excitations are generated by the action of the \Win\ in a
completely analogous manner. But now, if $m > 1$, the current
$j \equiv W^{(1)}$ contained in the \Win\ is not longer sufficient to
generate all the
edge excitations. The partition function (\ref{eq:part}) has to be replaced
by its $m^{\rm th}$ power, i.e.\ the edge excitations are
generated by $m$ currents \cite{Wen92}. In the same way it is possible
to reproduce the hierarchy picture of F.D.M.~Haldane and B.I.~Halperin
\cite{FZ91,WZ92}.

\mysection{Conclusion}
\pn
In this paper we have shown that the \Win\ is the underlying symmetry
in the IQHE as well as in the FQHE which generates all edge excitations.
This \Win\ was first introduced in the case $\nu = 1$ IQHE by
A.~Cappelli, C.A.~Trugenberger and G.R.~Zemba, describing the
incompressibility of the quantum droplet.
We have shown that the Laughlin wave functions for
$\nu={1\over 2p+1}$ can be interpreted as highest weight vectors of a
\Win\ which describes the quantized algebra of the area preserving
diffeomorphisms. For this
generalization we have introduced an electron-electron interaction
term which can be considered as adding flux quanta to each electron
as in J.K.~Jain's composite fermion theory. Further, we calculated all
edge exciations of this quantum droplet interpreting them as area
preserving surface deformations and we could show that these are
surface waves
which are moving with the same velocity and in the same direction.
\pn
There exit a lot of other examples of trial wave functions not only for
filling fraction $\nu={1 \over 2p+1}$. We have applied our methods
to wave functions for multi layer systems and systems of interacting
Landau levels.
\pn
An open question still is how the Coulomb interaction in the solid breaks
this symmetry.
\bigskip\pn{\bf Acknowledgements}\pn
We would like to thank W.~Nahm, A.~Honecker, M.~R\"osgen and M.~Terhoeven
for useful discussions and
careful reading of the manuscript. Moreover, we thank all participants
of the workshop
on the quantum Hall effect in Haus Windersch (M\"unster im Taunus, Germany).
\pn
This work is supported by the Deutsche Forschungsgemeinschaft.

\mysection{Appendix}
In this appendix we sketch a derivation of equation (\ref{eq:mammut}).
First, one shows
inductively that
  \begin{equation}
    (b_i)^n\psi_p = \sum_{{{1\leq j_k\neq j_l\leq N \atop j_k,j_l\neq
    i} \atop (1\leq k<l\leq n)}}
    \prod_{k=1}^{n}\frac{1}{z_i - z_{j_k}}\psi_p\,,
  \end{equation}
Thus, the action of $W_n^{(s)}$ on $\psi_p$ is given by
  \begin{equation}
    W_n^{(s)}\psi_p = \sum_{{1\leq j_k\neq j_l\leq N \atop (0\leq k<l
    \leq s-1)}}
    \prod_{k=1}^{s-1}\frac{z_{j_0}^{n+s-1}}{z_{j_0} - z_{j_k}}\psi_p\,.
  \end{equation}
Note, that this expression does not explicitly depend on $p$. Now, we can
rewrite the sums in terms of determinants in the following way:
  \begin{eqnarray}
    & &\sum_{{1\leq j_k\neq j_l\leq N \atop (0\leq k<l\leq s)}}
    \prod_{k=1}^{s}\frac{z_{j_0}^{r}}{z_{j_0} - z_{j_k}}\\
    &=&\sum_{1\leq j_0<j_1<\ldots<j_s\leq N}\sum_{\sigma\in S_{s+1}/S_{s}}
    |S_{s}|\frac{z_{\sigma(j_0)}^{r}}{z_{\sigma(j_0)} - z_{\sigma(j_k)}}\\
    &=&s!\sum_{1\leq j_0<j_1<\ldots<j_s\leq N}\sum_{l=0}^{s}(-)^{l+1}z_{j_l}^r
    \prod_{k=1}^{l-1}(z_{j_k}-z_{j_l})^{-1}
    \prod_{k=l+1}^{s}(z_{j_l}-z_{j_k})^{-1}\\
    &=&s! \sum_{1\leq j_0<j_1<\ldots<j_s\leq N}\frac{
    \sum_{l=0}^{s}(-)^{l+1}z_{j_l}^r\prod_{m<n\atop m,n\neq l}
    (z_{j_m}-z_{j_n})}{
    \prod_{i<k}(z_{j_i}-z_{j_k})}\,,
  \end{eqnarray}
where we sum over all fixed point free permutations of $S_{s+1}$ and
where the
sign comes from the asymmetry of the factors $(z_i - z_j)$. The last
expression is nothing but the expansion of a determinant divided by a
Vandermonde determinant, hence we arrive at eqn.\ (\ref{eq:mammut}).

\end{document}